\documentclass{emulateapj}
\usepackage[]{times, graphicx}
\newcommand{\alfven}{Alfv\'{e}n}
\newcommand{\alfvenic}{Alfv\'{e}nic}
\newcommand{\pref}{\protect\ref}
\begin{document}

\shorttitle{Coronal Dimming}
\shortauthors{S.W.~McIntosh} 
\title{The Inconvenient Truth About Coronal Dimmings} 
\author{Scott W. McIntosh} 
\affil{High Altitude Observatory,\\National Center for Atmospheric Research,\\P.O. Box 3000, Boulder, CO 80307}
\email{mscott@ucar.edu}

\begin{abstract}
We investigate the occurrence of a CME-driven coronal dimming using unique high resolution spectral images of the corona from the Hinode spacecraft. Over the course of the dimming event we observe the dynamic increase of non-thermal line broadening in the 195.12\AA{} emission line of \ion{Fe}{12} as the corona opens. As the corona begins to close, refill and brighten, we see a reduction of the non-thermal broadening towards the pre-eruption level. We propose that the dynamic evolution of non-thermal broadening is the result of the growth of \alfven{} wave amplitudes in the magnetically open rarefied dimming region, compared to the dense closed corona prior to the CME. We suggest, based on this proposition, that, as open magnetic regions, coronal dimmings must act just as coronal holes and be sources of the fast solar wind, but only temporarily. Further, we propose that such a rapid transition in the thermodynamics of the corona to a solar wind state may have an impulsive effect on the CME that initiates the observed dimming. This last point, if correct, poses a significant physical challenge to the sophistication of CME modeling and capturing the essence of the source region thermodynamics necessary to correctly ascertain CME propagation speeds, etc.
\end{abstract}


\section{Introduction}
Coronal dimmings, or ``transient coronal holes'' as they are sometimes known, have provoked great curiosity in the solar physics community since their initial observation with Skylab \citep[][]{Rust1976, Rust1983}. They were first noticed as a rapid intensity reduction of the soft X-Ray corona around active regions, and have subsequently been connected to coronal mass ejections \citep[CMEs; see, e.g.,][]{Forbes2000, Kahler2001}. Indeed, coronal dimmings are now viewed as the residual footprint of the CME in the corona \citep[e.g.,][]{Thompson2000}, the radio and plasma signatures of which are observed in interplanetary space \citep[e.g.,][]{Cane1984, Neugebauer1997, Attrill2008}. The connection between dimming and CME has been quantitatively fortified by the recent statistical surveys of \citet{Reinard2008} and \citet{Bewsher2008} using EUV instrumentation on the {\em SOHO} spacecraft \citep[][]{Fleck1995}. The former of these surveys indicates that at least 50\% of front-sided CMEs have associated dimming regions, while the latter stressed that the relationship between the two phenomena is one that grows considerably when only narrowband spectroscopic observations are considered. Therefore, rigorously establishing the poorly understood physical connection between CMEs and coronal dimmings using detailed spectroscopic measurement is a must.

We focus our analysis on observations of NOAA AR~10930 from the Extreme-ultraviolet Imaging Spectrometer \citep[EIS;][]{Culhane2007} on {\em Hinode} \citep[][]{Kosugi2007} between 19:00UT December 14 2006 and 06:00UT December 15 2006. This time period saw an X-Class flare and a $\sim$1000km/s halo CME\footnote{The CME properties were automatically derived from {\em SOHO}/LASCO data by the NASA/GSFC CDAW ({\url http://cdaw.gsfc.nasa.gov/}) and the Royal Observatory of Belgium/SIDC CACTUS \citep[][\-- {\url http://www.sidc.be/cactus/}]{Robbrecht2004} catalogues.} emanating from this complex active region at around 20:12~UT \citep[relative to SOHO/EIT imaging; ][]{Boudine1995}. In this Letter, we expand on the analysis of \citet{Harra2007}, exploiting rare detailed spectroscopic measurements of a dimming region. EIS provides a tantalizing look at the dynamic behavior of EUV emission lines, and their non-thermal line widths in particular, over the course of the eruption. The interpretation of the dynamic evolution of the non-thermal line widths presented in this Letter forms a challenge to the rapidly increasing sophistication of numerical CME models, in that they need to cope with the complex thermodynamics of the CME source region.

\begin{figure}
\epsscale{0.65}
\plotone{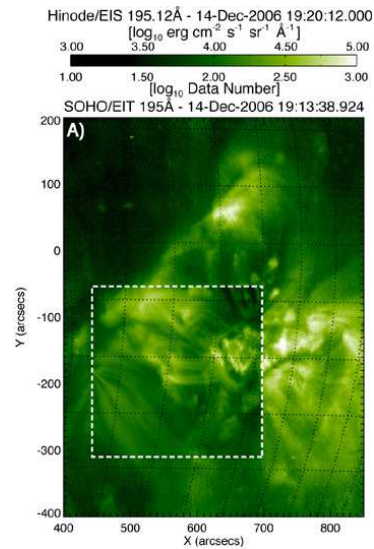}
\caption{Contextual pre-CME images of NOAA AR 10930 from {\em SOHO} and {\em Hinode}. We show the closest EIT 195\AA{} image to the start of the first EIS raster - shown inset as the peak intensity of the \ion{Fe}{12} 195.12\AA{} emission line. See the online edition of the journal to see a movie of the dimming evolution. \label{f1}}
\end{figure}

\section{Observations \& Data Analysis}\label{obs}
The dataset of interest comprises of three spectroheliogram ``raster'' observations (19:20-21:34UT, 01:15-03:30, 04:10-06:24UT), targeted at the following edge of the active region that is the source of the event studied. Each EIS raster is comprised of 256 horizontal (West to East) mirror mechanism steps with the 1\arcsec{} slit at a spacing of 1\arcsec{} and a height of 256\arcsec{} and has information in a nine 24 pixel wide spectral window. At a spectral resolution of 22.3m\AA{}, and wavelength of 195\AA{}, one pixel on the detector is equivalent to a velocity of $\sim$34~km/s.

For brevity, we present the analysis of the \ion{Fe}{12} 195.12\AA{} emission line as the EIS rasters can be supported by the broadband imaging from {\em SOHO}/EIT in the 195\AA{} passband at its regular 12 minute ``CME Watch'' cadence. The analysis of other strong and (relatively) spectrally clean emission lines observed in this period (\ion{Fe}{13} 202.04\AA{} and \ion{Fe}{15} 284.16\AA) will be published in a later article \citep[][]{McIntosh2009}.

Once the data are reduced, we need to extract the physical measurements from the emission line spectra. To do this, we choose to fit the spectral profiles at each spatial pixel at each raster step with a single Gaussian (assuming a linear continuum background variation in the spectral window) and, to ensure the highest quality fits, we use a combination of Genetic Algorithm \citep[e.g.,][]{McIntosh1998} and downhill convergence methods. Upon completion of the Gaussian fitting process, we have spatial maps of the line (peak) intensity, line position and $1/e$ width $w_{1/e}$ of the profile (measured in spectral pixels). The map of the non-thermal line width $v_{nt}$ (in km/s) of the emission line profile is determined using the quadratic relationship: 
\begin{equation}
v_{nt} = c \frac{D_{\lambda}}{\lambda_{0}} \sqrt { w_{1/e}^2 - w_{inst}^2 - w_{th}^2 },
\end{equation}
where $w_{inst}$ is the instrumental width \citep[taken to be 2.5 pixels full-width half-max;][]{Doschek2007}, $w_{th} (= \sqrt{2k_{B}T_{e}^{\ast} / m_{ion}})$ is the thermal width for an ion of mass $m_{ion}$ and peak formation temperature of $T_{e}^{\ast}$ (assuming that the ion and electron temperatures are equal), $\lambda_{0}$ (195.12\AA{}) is the rest wavelength of the line, and $D_{\lambda}$ (22.3m\AA{}) is the the spectral pixel scale and $c$ is the speed of light.

In order to compare the three EIS rasters, as well as with the broadband EIT images, we must de-rotate the coordinates of each slit position to the start-time of the first raster using the mapping method of \citet{McIntosh2006}. An example is shown in Fig.~\pref{f1}, where the monochromatic EIS image is inlaid in the closest EIT image to the start of the first raster (19:13UT).

\section{Results}\label{anal}
\begin{figure*}
\epsscale{1.}
\plotone{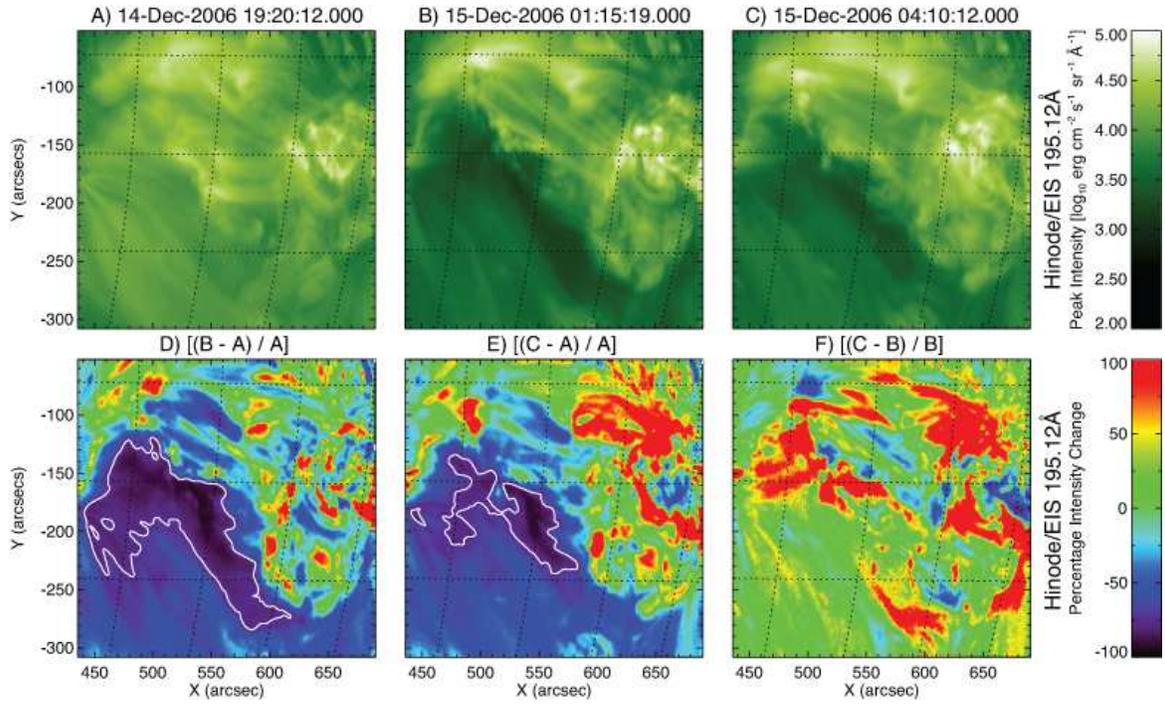}
\caption{Phases of the coronal dimming observed in the \ion{Fe}{12} 195.12\AA{} emission line intensity observed by EIS. Panels A through C show the line intensity before, at the peak and near the end of the dimming event, while panels D through F, respectively, show the percentage change in intensity between panels A and B, C and A, and C and B. The solid white contours in panels D and E indicate a 75\% reduction in intensity. \label{f2}}
\end{figure*}

Figure~\pref{f2} shows the evolution of the \ion{Fe}{12} line intensities before (19:20UT; panel A), and at two stages during (01:15; panel B -   04:10UT; panel C), the dimming event. Panel D shows the percentage change in the line intensity between the pre- and first post-eruption rasters [(B-A)/A], while panels E and F show the percentage intensity change between pre- and second post-eruption rasters [(C-A)/A] and that between the two post-eruption rasters [(C-B)/B]. The white contours overplotted on panels D and E isolate the strongest dimming region, where we see a reduction in intensity of 75\%. In fact, we see that, from EIS' perspective, there is a sizable reduction in intensity to the South and East of AR~10930; much of that area shows a drop in intensity between 25 and 50\%, which would classify this event as a ``deep'' dimming in the terminology of \citet{Attrill2007}. 

We should note here that the choice of a 75\% reduction in intensity is arbitrary, but that it outlines a region of significant change of the emission from magnetic field lines rooted in the region running from [440\arcsec, -200\arcsec] to [480\arcsec, -120\arcsec]. In panel F of Fig.~\pref{f2}, we see that the intensities are increasing over almost the entire region, recovering slowly as is typical of coronal dimmings \citep{Reinard2008,Attrill2008}. This behavior has been associated with the closure of the overlying corona and return of plasma heating to the region \citep[see, e.g.,][]{McIntosh2007}. We use the observed intensity drop to compute a rough estimate of the drop in electron density in the dimming region. For a resonance line whose intensity varies as $n_e^2$, a 75\% reduction in emission corresponds to a 44\% reduction of electron density. The apparent intensity drop and slow recovery of the region is mirrored by the {\em SOHO}/EIT image sequence (see the online edition of journal for movie accompanying Fig.~\pref{f1}). 

\begin{figure*}
\epsscale{1.}
\plotone{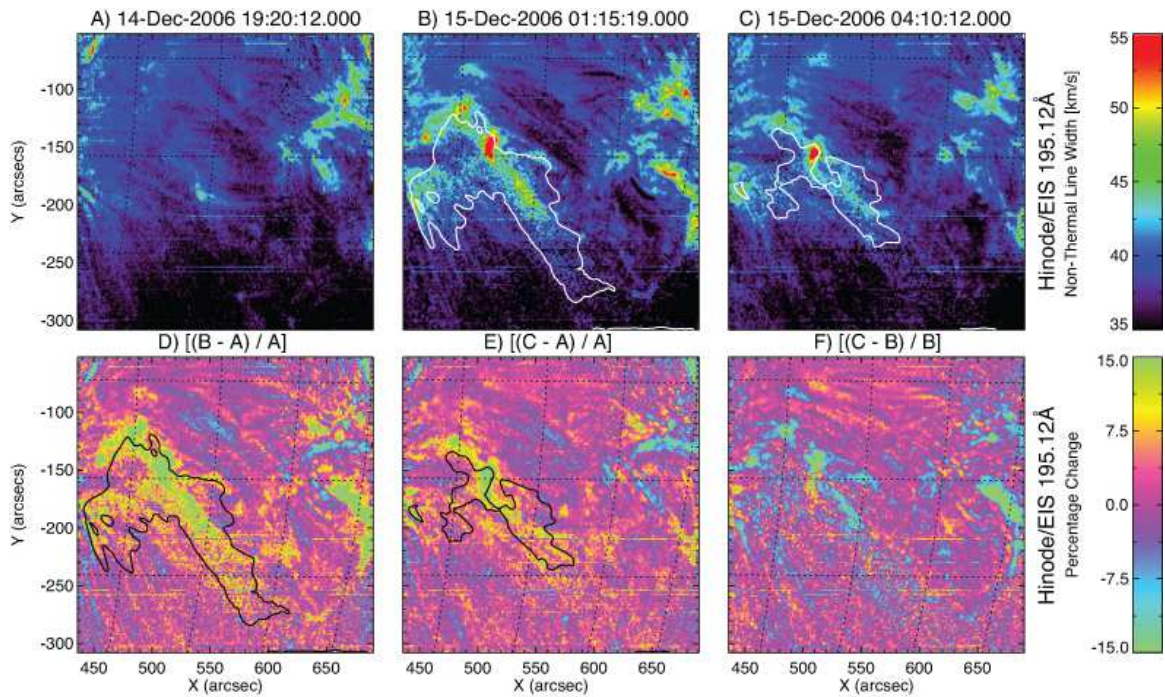}
\caption{Phases of the coronal dimming observed in the non-thermal line width of the \ion{Fe}{12} 195.12\AA{} emission line observed by EIS. Panels A through C show the line width before, at the peak and near the end of the dimming event, while panels D through F, respectively, show the percentage change in line width between panels A and B, C and A, and C and B. The solid white contours in panels B and C and the black contours in panels D and E indicate a 75\% reduction in intensity (from Fig.~\pref{f2}). \label{f3}}
\end{figure*}

Figure~\pref{f3} mirrors Fig.~\pref{f2}, except it shows the evolution of the \ion{Fe}{12} non-thermal line width ($v_{nt}$) before (panel A) and over the course of the dimming event (panels B and C). Again, panels D through F show the percentage change in $v_{nt}$ between the pre- and first post-eruption rasters [(B-A)/A], the pre- and second post-eruption rasters [(C-A)/A], and between the two post-eruption rasters [(C-B)/B]. Immediately, we see a dynamic evolution of $v_{nt}$ in the contour-outlined dimming region over the course of the event. Comparing panels A and B (and D), we see a large increase ($\sim$15\%) in $v_{nt}$, largely concentrated in and around the perimeter of the 75\% reduction contour. As the dimming event progresses (comparing panels C, A and E), there continue to be regions of difference in $v_{nt}$ late in the lifetime of the dimming, becoming more spatially compact and following the contraction of the intensity decrease contour. In panel F, we see that $v_{nt}$, much as in Fig.~\pref{f2}, appears to be slowly recovering to pre-eruption levels, and we find that other (spectrally clean) coronal spectral lines observed by EIS in this region exhibit the same dynamic behavior of $v_{nt}$.

\section{Discussion \& Conclusion}\label{discuss}
\citet{Harra2007} reported on the considerable (relative) blue-shifted Doppler velocities ($\sim$40km/s) that emanated from the dimming region and are indicative of significant coronal outflows while the region is open. We believe that these outflows are real, are consistent with the measurements discussed above, and reinforce our belief that the transient dimmings are really short lived coronal holes.  Unfortunately, the relative nature of the EUV Doppler measurements leaves some ambiguity in the absolute magnitude of these flows. However, we have demonstrated that the non-thermal line widths of these hot coronal emission lines also evolve dynamically, and are very responsive to the large-scale CME-related intensity fluctuations that historically constitute a coronal dimming event. 

The rapid growth of the coronal non-thermal line widths appears to be tied to the post-CME evacuation of the dimming region. Further, this rapid growth of non-thermal line widths is followed by a slow decrease, with values beginning to approach their pre-eruption levels - a result of the closing and gradual filling of the corona as the CME cuts its magnetic ties with the Sun \citep[][]{Reinard2008, Attrill2008}. 

In an effort to explain this evolution, we invoke the ubiquitous presence of \alfvenic{} plasma motions in the chromosphere \citep[e.g.,][]{DePontieu2007} and corona \citep[e.g.,][]{Tomczyk2007} and their likely connection to the non-thermal line widths measured in the upper atmosphere \citep[e.g.,][]{Tomczyk2007,McIntosh2008}. The observed increase in non-thermal line width in the dimming region is consistent with sub-resolution Doppler broadening resulting from the increased amplitude of \alfven{} waves in an increasingly rarefied plasma. For a constant volumetric energy flux in a uniform, unchanging, magnetic field, the amplitude of undamped \alfven{} waves should grow, as the density drops, by a factor of $\delta\rho^{-1/4}$. Further, as the corona begins to refill, the wave amplitudes should shrink back to their nominal (pre-event closed magnetic topology) value, precisely as we have observed in this instance. We stress that the dynamic behavior of the 195.12\AA{} non-thermal line widths is mimicked in the other spectrally isolated EIS lines studied for this event, but not shown in this Letter \citep[][]{McIntosh2009}.

The likely role of \alfven{} waves in the acceleration of the fast solar wind \citep[][]{Suzuki2005, DePontieu2007, Cranmer2007, Verdini2007} that originates in coronal holes and their observed connection following the passage of CMEs in interplanetary space \citep[e.g.,][]{Neugebauer1997} is evocative. If a coronal dimming is indeed a transiently evolving coronal hole, then it {\em must} carry some, if not all, of the same spectral, thermodynamic and compositional characteristics of its longer lived brethren. This implies that dimming regions are sources of fast wind streams blowing behind--in the magnetic envelopes of--CMEs. Such wind streams should ``switch-on'' on a timescale commensurate with the \alfven{} crossing time of the region and ``blow'' for a length of time commensurate with that required for the open magnetic flux behind the CME to close again. If this conjecture is indeed true, we would expect that the amount of unsigned magnetic flux in the dimming region should correlate strongly with the speed of the eventual Interplanetary CMEs seen in situ - scaling with the amount of open wave-carrying magnetic flux \citep[][]{Schwadron2008, McComas2008} - and observed by \citet{Chen2006}. 

The potential link between CMEs, coronal dimmings and induced fast solar wind streams will require an increase in the sophistication of current numerical CME models to capture the rapidly evolving complex thermodynamic state of the lower boundary. Only then will we be able to assess the impact of the following wind stream on the CME. Such measures are essential to correctly estimate the propagation speed of CMEs and necessary for accurate space weather predictions.

We surmise that the combination of the observed effect dictates that the evolution of the CME and the nearly instantaneous release of a fast wind stream are intrinsically coupled. This goes some way to explaining the frequent in situ observations of chromospheric compositional characteristics and the significant \alfvenic{} modulation that closely follows the CME in interplanetary space \citep[][]{Neugebauer1997, Skoug2004}. The influence of the dimming region-initiated wind stream on the flight characteristics of the CME remains to be seen. The hypothesis presented is one that we hope to directly test when the Coronal Multi-channel Polarimeter \citep[CoMP;][]{Tomczyk2007,Tomczyk2008} instrument begins regular observation from the Mees Solar Observatory on Haleakala early in 2009.

\acknowledgements
\begin{small}
SWM acknowledges support from NASA grants NNX08AL22G, NNX08AU30G  and fruitful discussions on this matter with Bart De Pontieu, Bob Leamon, David Alexander and Joan Burkepile who helped form the discussion presented. Further, SWM would like to thank Meredith Wills-Davey for very constructive comments made during the review process and for pointing out some relevant papers not known to the author. The National Center for Atmospheric Research is sponsored by the National Science Foundation. {\em SOHO} is a mission of international cooperation between ESA and NASA. {\em Hinode} is a Japanese mission developed and launched by ISAS/JAXA, collaborating with NAOJ as a domestic partner, NASA and STFC (UK) as international partners. Support for the post-launch operation of {\em Hinode} is provided by JAXA and NAOJ (Japan), STFC (U.K.), NASA, ESA, and NSC (Norway).
\end{small}

\end{document}